\documentclass[aps,prb,reprint,superscriptaddress,twocolumn]{revtex4}
\usepackage{units}
\usepackage{amsmath}
\usepackage{amssymb}
\usepackage{graphicx}
\usepackage{bm}
\usepackage{microtype,color}

\newcommand{\be}{\begin{equation}}
\newcommand{\ee}{\end{equation}}
\newcommand{\bea}{\begin{eqnarray}}
\newcommand{\eea}{\end{eqnarray}}

\begin{document}

\title{Control of Light Diffusion in a Disordered Photonic Waveguide}

\author{Raktim Sarma}
\affiliation{\textls[-20]{Department of Applied Physics, Yale University, New Haven, CT, 06520, USA}}
\author{Timofey Golubev}
\affiliation{\textls[-20]{Department of Physics, Missouri University of Science and Technology, Rolla, Missouri 65409,USA}}
\author{Alexey Yamilov}
\email{yamilov@mst.edu}
\affiliation{\textls[-20]{Department of Physics, Missouri University of Science and Technology, Rolla, Missouri 65409,USA}}
\author{Hui Cao}
\email{hui.cao@yale.edu}
\affiliation{\textls[-20]{Department of Applied Physics, Yale University, New Haven, CT, 06520, USA}}
\date{\today}

\begin{abstract}
We control the diffusion of light in a disordered photonic waveguide by modulating the waveguide geometry. In a single waveguide of varying cross-section, the diffusion coefficient changes spatially in two dimensions due to localization effects. The intensity distribution inside the waveguide agrees to the prediction of the self-consistent theory of localization. Our work shows that wave diffusion can be efficiently manipulated without modifying the structural disorder.
\end{abstract}


\maketitle

The concept of diffusion is widely used to study the propagation of light through multiple scattering media such as clouds, colloidal solutions, paint, and biological tissue \cite{1978_Ishimaru,Sheng1,Rossum1,Akkermanbook}. The diffusion, however, is an approximation as it neglects wave interference effects \cite{Sheng2}. Most of the scattered waves go through independent paths and have uncorrelated phases, so their interference is averaged out. However, a wave may return to a position it has previously visited after multiple scattering, and there always exists the time-reversed path which yields identical phase delay. Constructive interference between the waves traveling in the time-reversed paths increases the energy density at the original position, thus suppressing diffusion \cite{1979_Gorkov} and eventually leading to localization \cite{1958_Anderson}. This effect has been accounted for by a renormalized diffusion coefficient $D$ in the self-consistent theory of localization \cite{1980_Vollhardt_Wolfle,1993_Kroha_self_consistent}. The amount of renormalization depends on the return probability, which is determined by the size of a random medium as well as the position inside \cite{2000_van_Tiggelen,2008_Cherroret,2008_Tian,2010_Payne}. We recently reported a direct observation of the position-dependent diffusion coefficient in disordered waveguides \cite{Dz}. By changing the waveguide length and width, we tuned the diffusion coefficient by varying the strength of wave interference. However, the width of each waveguide was kept constant, and we switched between the waveguides to control diffusion.



In this Letter, we fabricate disordered waveguides with a variable cross-section and thus achieve control of light transport in the same system.
These new structures make it necessary to account for spatial variation of diffusion coefficient $D$ in two dimensions (2D) due to the modulation of the waveguide width. Experimentally we fabricate a random array of air holes in a waveguide geometry on a silicon wafer, and probe light propagation inside the 2D structure from the third dimension. The measured spatial distribution of light intensity inside the disordered waveguide agrees well to the prediction of the self-consistent theory of localization\cite{2008_Cherroret,2010_Payne}. Instead of changing the degree of disorder, we demonstrate that the wave diffusion can be manipulated by changing geometry (cross-section) of the random waveguide nano-structures.

\begin{figure}[htbp]
\centering
\includegraphics[width=0.8\linewidth]
{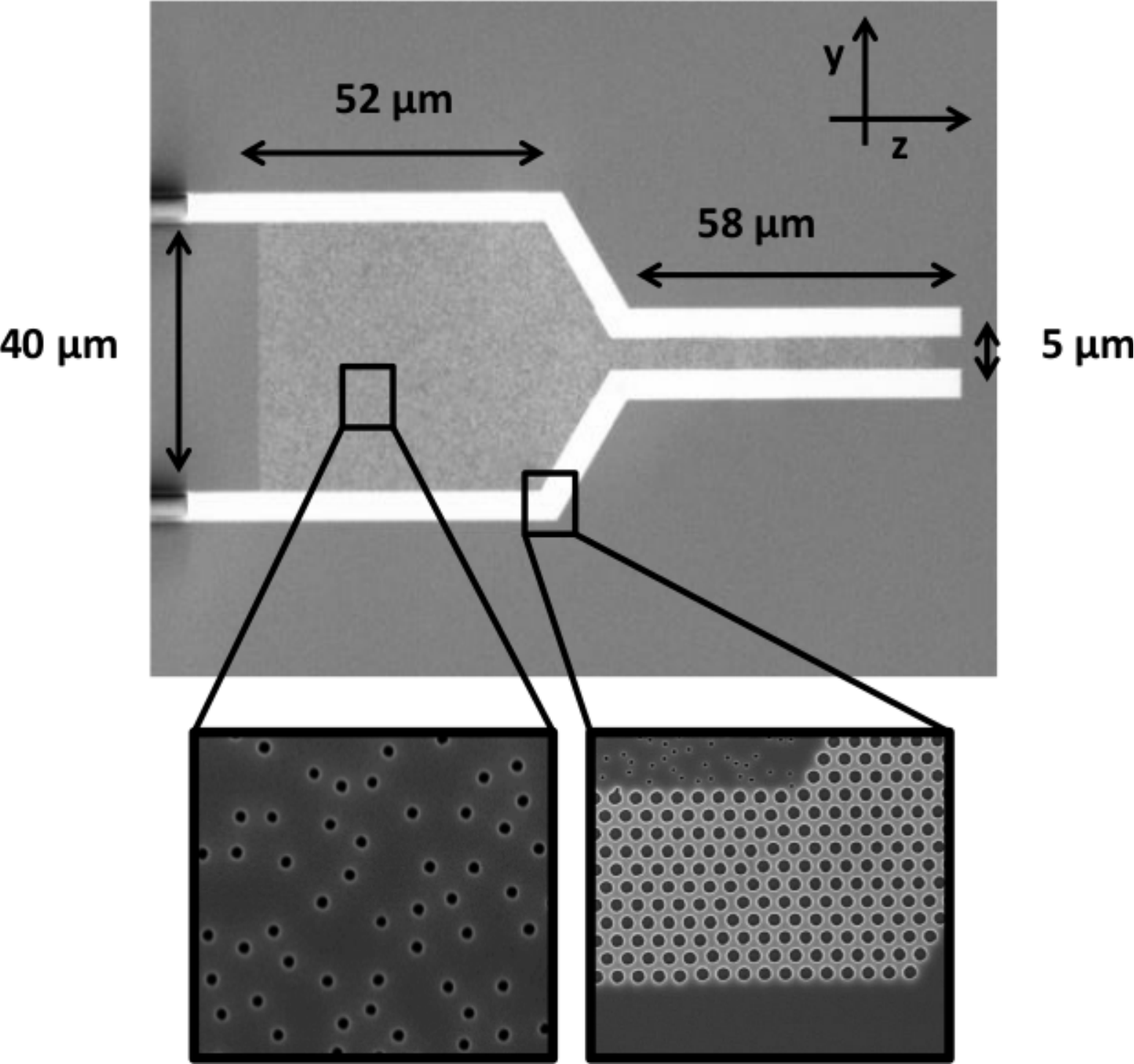}
\caption{Top-view scanning electron microscope (SEM) image of a quasi-2D disordered photonic waveguide. Light is injected from the left end of the waveguide and incident onto the random array of air holes. The waveguide wall is made of a triangle lattice of air holes which forms a 2D photonic bandgap to confine light inside the waveguide. The width of the random waveguide is changed gradually from 40 $\mu$m to 5 $\mu$m through a tapered region. }
\end{figure}

The disordered waveguides were fabricated with a silicon-on-insulator (SOI) wafer where the thickness of silicon layer and the buried oxide were 220 nm and 3$\mu$m respectively. The patterns were written by electron beam lithography and etched in an inductively-coupled-plasma (ICP) reactive-ion-etcher (RIE). Figure 1 is the scanning electron microscope (SEM) image of a fabricated sample. The waveguide contained a 2D random array of air holes. The hole diameters were 120 nm, and the average (center-to-center) distance of neighboring holes was about 385 nm. The total length $L$ of the random waveguide was 120 $\mu$m, and the waveguide width was changed from $W_1$ = 40 $\mu$m to $W_2$ = 5 $\mu$m via a tapered region. The lengths of wider ($W_1$) and narrower ($W_2$) sections were $L_1$ = 52 $\mu$m and $L_2$ = 58 $\mu$m respectively. The tapered section was 10 $\mu$m long, with a tapering angle of 60 degrees. The waveguide walls were made of triangle lattice of air holes (lattice constant 440 nm, hole radius 154 nm) that had complete 2D photonic bandgap.

In the optical experiment, we used a lensed fiber to couple monochromatic light (wavelength $\sim$ 1500 nm) from a tunable CW laser source (HP 8168F) into the waveguide [Fig. 2(a)]. The polarization of input light was transverse-electric (TE) (electric field parallel to the waveguide plane). Light was scattered by the air holes inside the waveguide and undergoes diffusion. The waveguide walls provided in-plane confinement of the scattered light. However, some of the light was scattered out of the waveguide plane. This leakage allowed us to observe light propagation inside the disordered waveguide from the vertical direction. The spatial distribution of light intensity across the waveguide was projected by a 50$\times$ objective lens [numerical aperture (NA) = 0.42] onto an InGaAs camera (Xeva 1.7-320). Figure 2(b) shows a typical near-field image, from which we extracted the 2D intensity distribution inside the waveguide $I(y,z)$.
The ensemble averaging was done over three random configurations of air holes and 25 input wavelengths equally spaced between 1500 nm and 1510 nm. The wavelength spacing was chosen to produce different intensity distributions. Further averaging was done by slightly moving the input beam position along the transverse $y$ direction. Nevertheless, the front surface of the random structures was always uniformly illuminated by the incident light.

\begin{figure}[htbp]
\centering
\includegraphics[width=1\linewidth]
{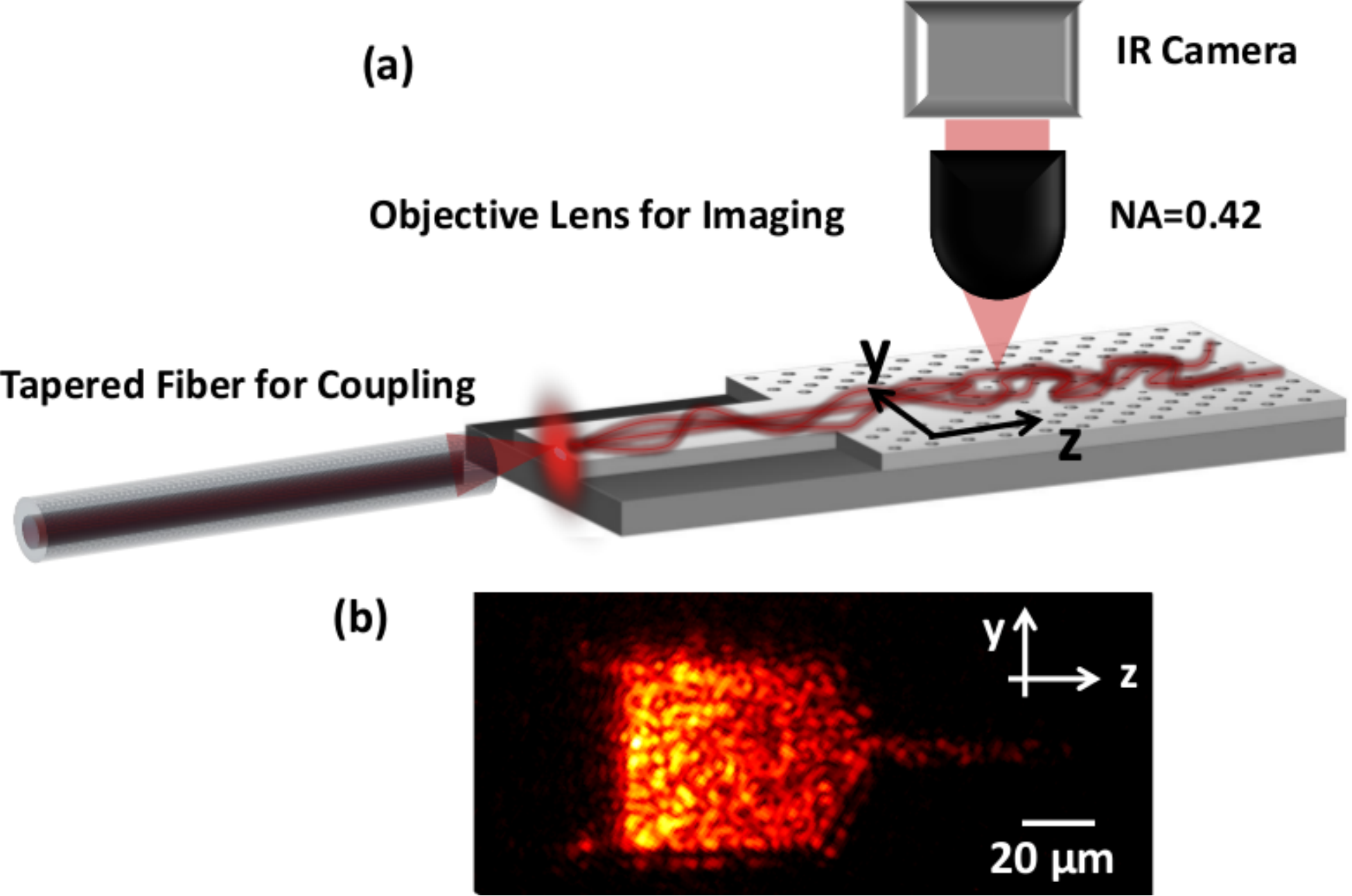}
\caption{(Color online) (a) A schematic of the experimental setup. A lensed fiber couples the light to the structure and another 50$\times$ objective lens (NA = 0.42) collects the light scattered by the air holes out of the waveguide plane and projects onto a camera. (b) Near-field optical image of the intensity of light scattered out-of-plane from the disordered waveguide. The wavelength of the probe light is 1505 nm.}
\end{figure}

The relevant parameters for light propagation in the disordered waveguide are the transport mean free path $\ell$ and the diffusive dissipation length $\xi_a$. The transport mean free path $\ell$ depends on the size and density of the air holes. The dissipation mostly comes from out-of-plane scattering as the silicon absorption at the probe wavelength is negligible. As shown in our previous work \cite{Dz}, this vertical loss of light can be treated a dissipation (as absorption) and described by the characteristic length $\xi_a = \sqrt{D_0 \tau_a}$, where $\tau_a $ is the ballistic dissipation time and $D_0$ is the diffusion coefficient without localization corrections.

There are three main advantages of using the planar waveguide geometry. First, it allows a precise fabrication of the designed structure so that the parameters such as the transport mean free path can be accurately controlled. Second, we can easily monitor the in-plane diffusion by collecting the out-of-the-plane scattered light.  Third, the localization length $\xi$ can be tuned by changing the waveguide width $W$, because $\xi= (\pi/2)N\ell$, where $N = 2W/(\lambda/n_e)$ is the number of propagating modes in the waveguide, which is proportional to $W$. By varying the width of a single waveguide, we adjust the strength of localization effect along the waveguide. The localization length in the wider section of the waveguide ($W_1$ = 40 $\mu$m) is 8 times longer than that in the narrower section ($W_2$ = 5 $\mu$m). Hence, the suppression of diffusion by wave interference is enhanced approximately 8 times in the narrower section of the waveguide.

For a quantitative description of light transport in the random waveguide of variable width, we used the self-consistent theory of localization to calculate the diffusion coefficient $D(y,z)$ inside the waveguide. The renormalization of $D$ depends on the return probability, which is position dependent \cite{2000_van_Tiggelen,2008_Cherroret,2008_Tian}. The maximum renormalization happens inside the random media at a location where the return probability is the highest, and the renormalization is lowest near the {\it open} boundaries of the random media. As it will be shown below, the return probability takes the maximum value in the narrow portion of the structure and not at the geometrical center as in waveguides with a uniform cross-section. The renormalization of the diffusion coefficient also depends on the amount of dissipation, which suppresses feedback from long propagation paths and sets an effective system size beyond which the wave will not return\cite{1998_Brouwer, 2013_Yamilov_Localization_with_Absorption}.

Numerically we computed $D(y,z)$ using the commercial package Comsol Multiphysics after setting the values of the transport mean free path $\ell$ and the diffusive dissipation length $\xi_a$. First the return probability was calculated at every point in the waveguide \cite{2010_Payne}. This was done by moving a point source throughout the structure and calculating the light intensity at the source for each source position. This intensity was taken as the return probability which was then used to renormalize $D(y,z)$. The modified $D(y,z)$ was then used to recalculate the return probability. Several iterations of this procedure were performed until the changes in $D(y,z)$ between iterations became small enough to be negligible. Once we obtained the final value of $D(y,z)$, it was used to calculate the intensity $I(y,z)$ inside the waveguide. The above calculation was repeated for various combinations of $\ell$ and $\xi_a$ until the calculated $I(y,z)$ matched the measured intensity distribution. The parameters that gave the best agreement were $\ell$ = 2.9 $\mu$m and $\xi_a$ = 35 $\mu$m. Figure 3(a) plots the calculated return probability, which is greatly enhanced by the stronger transverse confinement (along $y$ direction) in the narrower section of the waveguide. Consequently, the renormalized diffusion coefficient $D(y,z)$, shown in Fig. 3(b), reaches the minimum value close to the middle of the narrower section. Note that in the tapered region, $D$ changes not only along $z$, but also along $y$. The smaller $D$ near the boundary is attributed to the enhancement of return probability due to reflection from the photonic crystal wall. Figure 3(c) shows the spatial distribution of in-plane diffusive light intensity $I(y,z)$ inside the waveguide.

\begin{figure}[htbp]
\centering
\includegraphics[width=0.5\linewidth]
{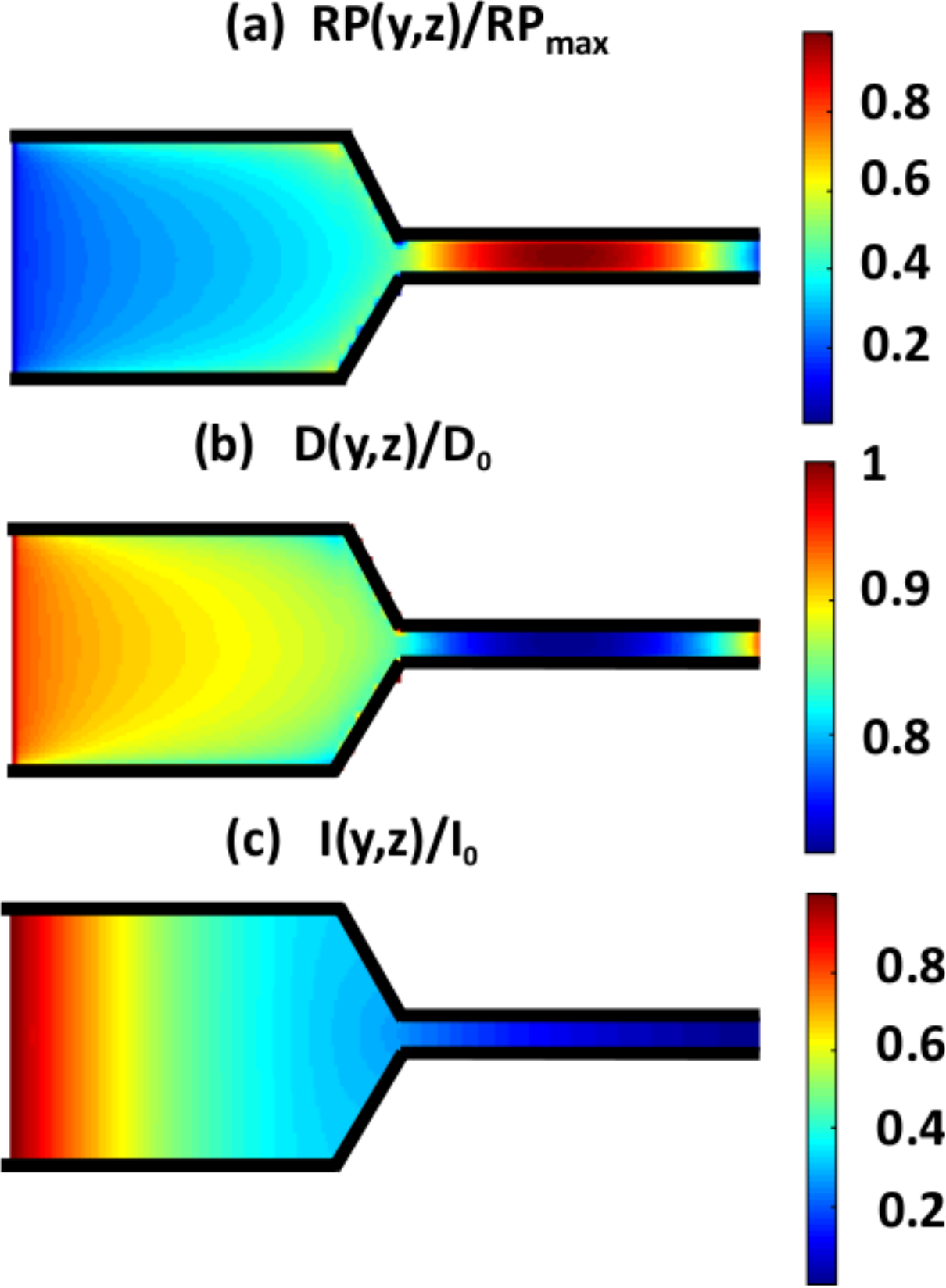}
\caption{(Color online) (a) Calculated return probability in the disordered waveguide shown in Fig. 1. $\ell=2.9\mu$m, and $\xi_a=35\mu$m. (b) 2D renormalized position dependent diffusion coefficient $D(y,z)/D_0$ for the same structure as in (a). (c) Intensity distribution $I(y,z)/I_0$ inside the random structure obtained from $D(y,z)/D_0$ in (b).
}
\end{figure}

\begin{figure}[htbp]
\centering
\includegraphics[width=1\linewidth]
{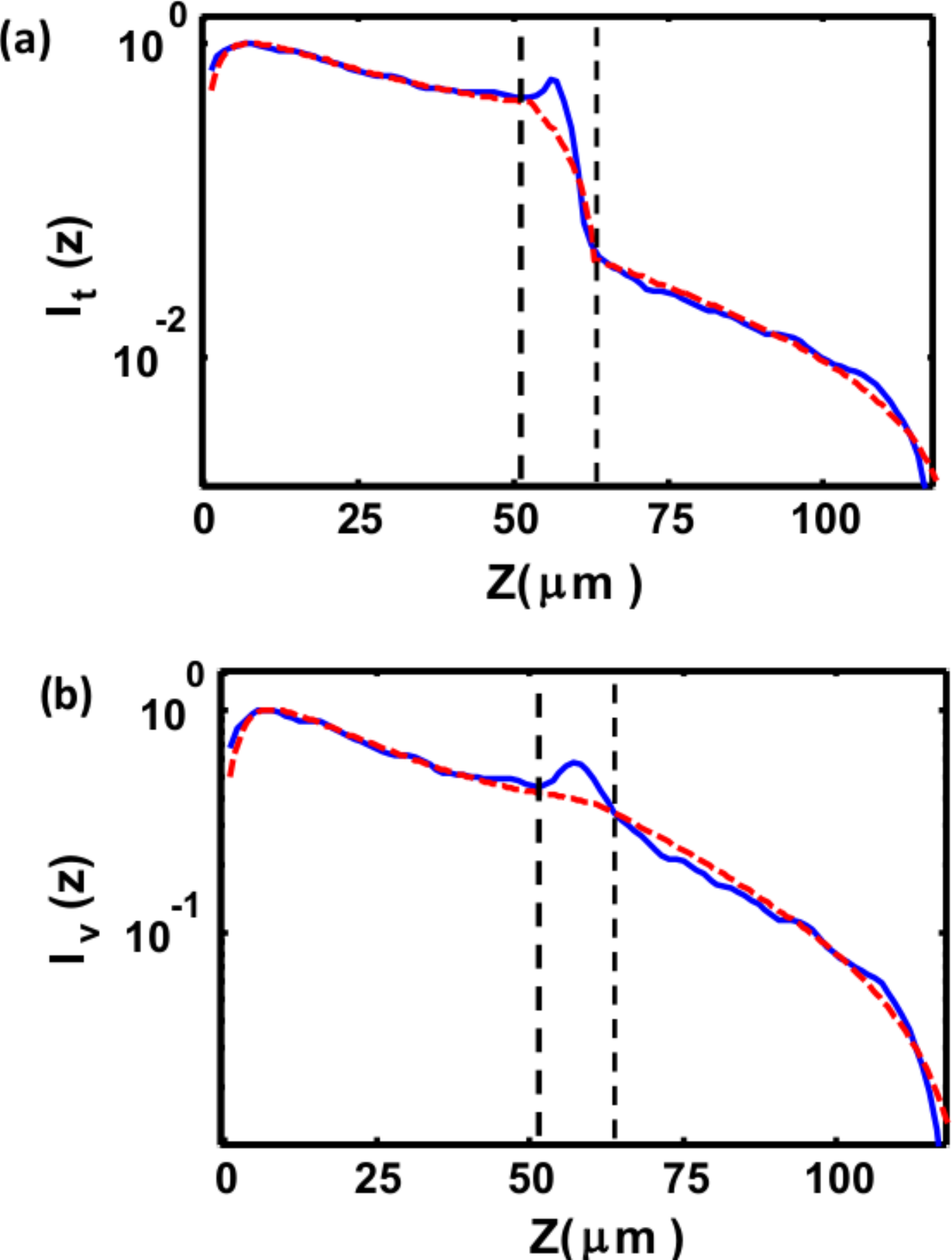}
\caption{(Color online) (a) Comparison of the measured cross-section integrated intensity  $I_t(z)$ (solid blue line) to numerical simulations (dashed red line). (b) Measured cross-section averaged intensity  $I_v(z)$ (solid blue line) in comparison with the calculated one (dashed red line). The vertical dotted lines in (a,b) marks the starting point and the end point of the tapered region.}
\end{figure}

From the measured $I(y,z)$, we computed the cross-section integrated intensity $I_t(z) = \int_{-W(z)/2}^{W(z)/2} I(y,z) dy$ and the cross-section averaged intensity $I_v(z) = I_t (z)/W(z)$. The former quantity is proportional to the $z$-component of the energy flux through the cross-section of the waveguide, while the latter quantity, $I_v(z)$, is related to the energy density. As shown in Fig. 4(a), $I_t(z)$ decays more slowly with $z$ in the wider section of the waveguide than in the narrower one.
This is attributed to two factors, (i) reflection from the boundary of the tapered region, (ii) stronger localization effect in the narrower section of waveguide. The narrowing of the waveguide width leads to a sharp drop of $I_t$ (energy flux), as part of the diffusive light is reflected back. The dashed curve in Fig.~4(a) is the calculated $I_t(z)$, which agrees well to the experimental data. Figure~4(b) plots the measured $I_v(z)$ together with the calculated one. Again we see a good agreement except at $z\sim60$ $\mu$m. The near-field optical image [Fig. 2(b)] reveals that near the photonic crystal wall of the tapered section, the abrupt backward scattering leads to formation of standing wave, thus the intensity is enhanced compared to the diffusive prediction. The spatial extent of this effect is determined by the transport mean free path $\ell$ beyond which the direction of the reflected wave is randomized. Inherent inability of a diffusive description to describe transport on scales shorter than $\ell$ explains the deviation of the experimentally measured intensity from the theoretical prediction as exhibited in Fig. 4(b) by a small bump at $z\sim60$ $\mu$m.


In conclusion, we demonstrated an effective way of manipulating light diffusion in a disordered photonic waveguide. Instead of changing the degree of structural disorder, we varied the waveguide geometry (its cross-section). By modulating the width in a single waveguide, we manipulated the interference of scattered light and made the diffusion coefficient vary spatially in two dimensions. We measured the intensity distribution inside the quasi-2D random structures by probing from the third dimension and the experimental results agreed well to the predictions of the self-consistent theory of localization. Although, the experiments in this work were done with light, the outlined approach to control diffusion is also applicable to other types of waves, such as acoustic wave, microwave and the de Broglie wave of electrons.

We acknowledge Douglas Stone and Arthur Goetschy for useful discussions and suggestions. We also thank Michael Rooks for suggestions regarding fabrication of the sample. This work was supported by the National Science Foundation under grants nos. DMR-1205307, DMR-1205223 and ECCS-1128542. Facilities used were supported by YINQE and NSF MRSEC Grant No. DMR-1119826.


\begin{thebibliography}{99}


\bibitem{Sheng1}{\it Scattering and Localization of Classical Waves in Random Media},edited by P. Sheng (World Scientific, Singapore,1990).

\bibitem{1978_Ishimaru} A. Ishimaru, {\it Wave Propagation and Scattering in Random
  Media}(Academic Press,1978).

\bibitem{Rossum1} M. C. van Rossum \& T. M. Nieuwenhuizen, {\it Rev. Mod. Phys.} {\bf 71}, 313 (1999).

\bibitem{Akkermanbook} E. Akkermans \& G. Montambaux,{\it Mesoscopic Physics of Electrons and Photons}(Cambridge University Press, Cambridge,2007).

\bibitem{Sheng2}Ping Sheng, {\it Introduction to Wave Scattering, Localization, and Mesoscopic Phenomena}(Academic,Boston,1995).

\bibitem{1979_Gorkov} L. Gor'kov et al., {\it JETP Lett.} {\bf 30}, 228 (1979).

\bibitem{1958_Anderson} P. W. Anderson, {\it Phys. Rev.} {\bf 109}, 1492 (1958).

\bibitem{1980_Vollhardt_Wolfle}D. Vollhardt \& P.W\"olfle, {\it Phys. Rev. B} {\bf 22}, 4666 (1980).

\bibitem{1993_Kroha_self_consistent}J. Kroha et al., {\it Phys. Rev. B} {\bf 47}, 11093 (1993).




\bibitem{2000_van_Tiggelen}B. A. van Tiggelen et al., {\it Phys. Rev. Lett.} {\bf 84}, 4333 (2000).

\bibitem{2008_Cherroret}N. Cherroret \& S.E Skipetrov, {\it Phys. Rev. E} {\bf 77}, 046608 (2008).

\bibitem{2008_Tian}C. Tian, {\it Phys. Rev. B} {\bf 77}, 064205 (2008).

\bibitem{2010_Payne}B.Payne, A.Yamilov, \& S.E.Skipetrov, {\it Phys. Rev. B} {\bf 82}, 024205 (2010).

\bibitem{Dz}A. Yamilov et al., {\it Phys. Rev. Lett.} {\bf 112}, 023904 (2014).

\bibitem{1998_Brouwer}P. W. Brouwer, {\it Phys. Rev. B} {\bf 57}, 10526 (1998).

\bibitem{2013_Yamilov_Localization_with_Absorption} A. Yamilov \& B. Payne, {\it Optics Express} {\bf 21}, 11688 (2013).

\end{thebibliography}
\end{document}